\documentclass[,nofootinbib,aps,prl,showpacs,preprintnumbers,
amsmath,amssymb,superscriptaddress,twocolumn]{revtex4}

\usepackage{graphicx}

\begin{document}


\title{ On the Sudakov suppression and enhancement for 
electroweak reactions }

\vspace*{0.3 cm}

\author{B.I.~Ermolaev}
\affiliation{CFTC, University of Lisbon
Av. Prof. Gama Pinto 2, P-1649-003 Lisbon, Portugal \\
and \\
Ioffe Physico-Technical Institute, 194021
 St.Petersburg, Russia}

\begin{abstract}
Accounting for the double-logarithmic (DL) contributions to amplitude of  
$Z \to f\,\bar{f}$ leads to the Sudakov suppression. In contrast,   
resumming DL contributions to amplitudes of  $W^{\pm} \to f\,\bar{f}'$ 
results into an exponential with a positive exponent, 
i.e. into the enhancement. This effect is intrinsic property of the theories 
with non-Abelian gauge groups.  
  
\end{abstract}

\pacs{12.38.Cy}

\maketitle

\section{Introduction}
\label{INTRO}

Among different double-logarithmic (DL) 
contributions, there are the contributions that  
depend on the infrared cut-off. Results of resuming them to all orders in the 
coupling(s) can be often written in an exponential form, with a negative 
exponent, so that it falls 
when the total energy grows. This is known as the Sudakov 
suppression~\cite{sud}. It means  the  
suppression of cross-sections of 
non-radiative processes compared to cross sections of 
the processes where bremsstrahlung is allowed. Such a suppression in QED 
(see e.g review \cite{g}) and in  QCD\cite{quarkff} 
was studied in details to 
all orders in the couplings. 
Exponentiation of the first-loop electroweak (EW) contributions 
was obtained in Ref.~\cite{flmm}  
and then was confirmed in Refs.~\cite{kp}-\cite{bw} by 
the fixed orders calculations. The complete list of publications on 
this subject can been found in the recent Refs.~\cite{ba},\cite{m}.  

However, resumming DL contributions of the same infrared origin sometimes  
leads to different results. In particular,  
it was found in Ref.~\cite{hol} that one-loop DL contribution to the 
$W^{\pm} \to f\,\bar{f'}$ has the opposite sign compared to the one for  
$\gamma \to  f\,\bar{f}$ and $Z \to f\,\bar{f}$ -decays. It suggests 
that the total resummation of the DL corrections to the amplitude of 
the $W^{\pm} \to f\,\bar{f'}$ may be an exponential with a positive 
exponent.   
   
In the present paper we 
consider amplitudes for the $W$ -decay. We show that the result of 
accounting for DL contributions to all orders 
in the EW couplings has an exponential form with the 
positive exponent. In other words, there is the enhancement instead of 
the suppression for such amplitudes. We consider the  
decays of the electroweak bosons into the left fermions $f$ and the right 
antifermions $\bar{f},~\bar{f'}$ though drop the subscripts $L$ and $R$ 
through the paper.     

The paper is organised as follows: in Sect.~II we briefly review 
the Sudakov suppression for amplitude of $Z \to f\bar{f}$, comparing the  
use the Feynman and the Coulomb gauges.  In Sect.~III we 
discuss the decay of $W$ -bosons into fermions, again with using both the 
Feynman and the Coulomb gauges, and arrive at the 
enhancement. The origin of the enhancement is considered in  in 
Sect.~IV.  
Finally, Sect.~V is for concluding 
remarks.            

\section{Sudakov suppression }
\label{SUP}

The Sudakov suppression\cite{sud} is well-known, so we 
discuss it very briefly. 
Let us consider amplitude $A_Z$ of the decay $Z \to f \bar{f}$ where $f$ 
stands for any (charged) fermion from the left doublets of the 
Standard Model. As we assume all external particles to be on-shell, $A_Z$ is 
gauge-independent. Among the often used gauges for propagators 
$D_{\mu \nu}(k)$ of the EW bosons, there  
are the Feynman gauge  and the 
Coulomb (temporal) gauge.  

\begin{figure}[htbp]
  \begin{center}
    \includegraphics[scale=.7]{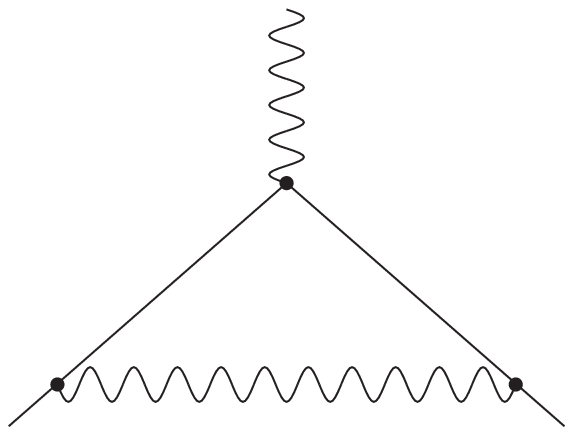}
    \caption{}
    \label{fig.1}
  \end{center}
\end{figure}

 In the first-loop approximation and 
when the Feynman gauge is used, the DL contribution to $A_Z$ comes from the 
graph in Fig.~1. The vertical waved line in Fig.~1 denotes the  
external EW boson. The horizontal waved line may correspond to any 
virtual electroweak boson though the DL contribution to  $A_Z$ appears    
only when this boson  
is a photon. When the Coulomb 
(temporal) gauge is used, DL contributions can come from the 
self-energy graphs depicted in Fig.~2. Again, DL contributions come  
only when the virtual bosons in Fig.~2 are photons. 
    
\begin{figure}[htbp]
  \begin{center}
    \includegraphics[scale=.7]{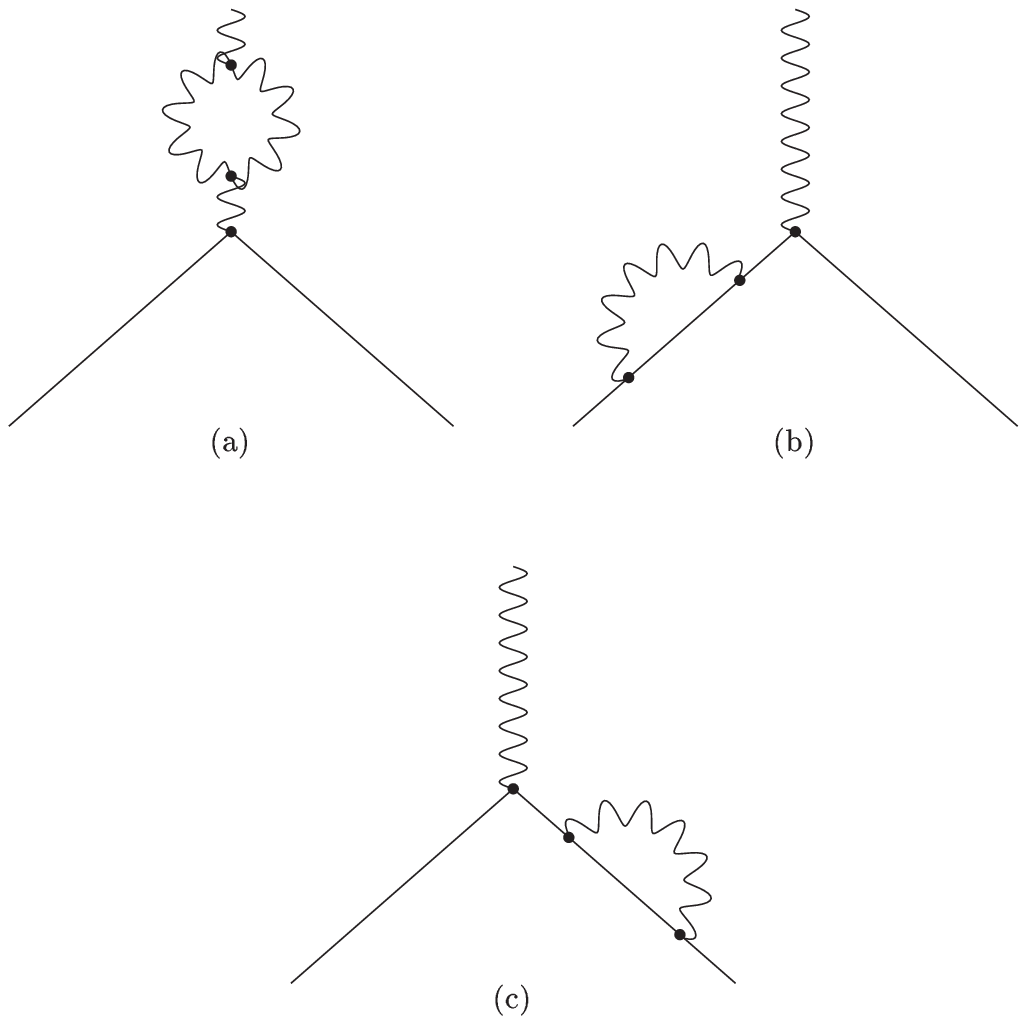}
    \caption{}
    \label{fig.2}
  \end{center}
\end{figure}

Using the Feynman 
gauge, it is easy to arrive at the well-known one-loop DL contribution 
$A_Z^{(1)}$ to $A_Z$ :

\begin{equation}
\label{rz}
A_Z^{(1)}/ A_Z^{Born}  = 
- Q^2_f \Big(\frac{\alpha}{4 \pi} \ln^2(-M^2/\mu^2) \Big) \equiv -Q^2_f L  
\end{equation}
where $Q_f$ is the electric charge of the fermion $f$, $M$ is the mass of 
the $Z$ -boson and $\mu$ is the infrared cut-off. With DL accuracy, one can 
neglect the difference between masses of $Z$ and $W$ -bosons.  
The double logarithm in Eq.~(\ref{rz}) is written in the simplest way 
for the particular case when $\mu \geq m$, with $m$ being the mass of the 
fermion $f$.
It also easy to arrive at Eq.~(\ref{rz}) when the Coulomb gauge 
 is used. In this case, DL contributions come from the 
self-energy graphs depicted in Fi.~2. The DL contributions  
$\Sigma_f^{DL}$ from the fermion self-energy graphs (graphs (b) and 
(c) in Fig.~2) are  
$\Sigma_f^{DL} = - Q^2_f L/2$.      
In order to get a DL contribution from graph~(a) in Fig.~2, at least one of 
the virtual bosons in Fig.~2a 
has to be a photon. However, there is not such a vertex 
in the Lagrangian of the Standard Model. It leaves us with 
two  contributions $ \Sigma_f^{DL}$ and we are back to  
Eq.~(\ref{rz}). Using the Coulomb gauge for the Sudakov logarithms 
was discussed in details in  
Refs.~\cite{sv},\cite{bw}.   
The DL asymptotics of $A_Z$ accounting for contributions both  
$\sim \gamma_{\mu}$  
and $\sim \sigma_{\mu \nu} = 
(1/2)\left(\gamma_{\mu}\gamma_{\nu} - \gamma_{\nu}\gamma_{\mu} \right)$ 
can be obtained by direct graph-by-graph calculations like it  
was done  
 in  Ref.~\cite{et}. The complete expression for  the 
DL asymptotics for  
for $A_Z$ is    

\begin{equation}
\label{zdecay}
A_Z =  \Big[ \gamma_{\mu} + 
\frac{\sigma_{\mu \mu} q_{\nu}}{m} \frac{\partial}{\partial \rho}\Big]
e^{- Q^2_f (\alpha/4\pi) \ln^2\rho }  
\end{equation}
where we have omitted the spinors of the fermions and 
the $Z$-boson polarization vector,  
 $q$ is the sum of the fermion momenta  
$(q^2 = M^2)$;  
 we have chosen in Eq.~(\ref{zdecay}) $\mu \approx m$ 
and denoted $\rho \equiv - M^2/m^2$. 
The minus sign in the exponential of  Eq.~(\ref{zdecay}) means the Sudakov 
suppression. It suppresses the non-radiative decays of $Z$ -bosons.
    
\section{Sudakov enhancement}
\label{ENC}

Now let us consider DL contributions to amplitude $A_W$ of the  
decay $W^{\pm} \to f \bar{f}'$ where   
$f$ and $f'$ $(f \neq f')$  belong the same doublet of the Standard Model.   
When the Feynman gauge is used, DL 
contribution comes from the graph in Fig.~1 where the horizontal waved line 
denotes again the virtual photon. This contribution is   

\begin{equation}
\label{woneloop}
A_W^{(1)}/ A_W^{(Born)} = - Q_f Q_{f'} L  
\end{equation} 
where  $Q_f (Q_{f'})$ is the electric charge of the fermion $f ~(f')$. 
Invoking results of Ref.~\cite{et} leads immediately to the DL 
asymptotics for $A_W$: 

\begin{equation}
\label{wdecay}
A_W =  \Big[ \gamma_{\mu} + 
\frac{\sigma_{\mu \mu} q_{\nu}}{m} \frac{\partial}{\partial \rho}\Big]
e^{- Q_fQ_{f'} (\alpha/4\pi) \ln^2\rho } .  
\end{equation}
The term $\sim \alpha$ in the expansion  of Eq.~(\ref{wdecay}) into series in 
$\alpha$  was obtained in Ref.~\cite{hol}.  

Let us notice that when the fermions  
$f$ and $f'$ in  Eq.~(\ref{wdecay}) are quarks, they 
have opposite signs. It means that 
the exponent in  Eq.~(\ref{wdecay})  is positive  
and therefore  
there is the DL enhancement in  Eq.~(\ref{wdecay})  instead of the DL 
suppress of  Eq.~(\ref{zdecay}). For reactions    
$W^+ \to u \,\bar{d}$ and $W^- \to d \,\bar{u}$, the DL amplitude of 
the process is  
 
\begin{equation}
\label{wdecayq}
A_W(quarks) = A_W^{(0)}(quarks) e^{(2/9) L} ~. 
\end{equation} 
Notation $A_W^{(0)}(quarks)$ in  Eq.~(\ref{wdecayq}) stands for the 
expression in the squared brackets in  Eq.~(\ref{wdecay}) .
At the same time,   
Eq.~(\ref{wdecay}) reads that there are no 
DL contributions to the lepton decays $W \to l \bar{\nu}_l, 
W \to \nu_l \bar{l}$.   
Although amplitudes  $A_W$ are    
gauge-invariant, it is useful to demonstrate how the first-loop contribution 
$A_W^{(1)}$ of  Eq.~(\ref{woneloop}) can be easily obtained  
when the Coulomb gauge is used. 
In this case, DL contributions come from the self-energy 
graphs. Contributions of graphs (b) and (c) in Fig.~2 are 
 $\Sigma_f^{DL} = - Q^2_f L/2$  and $\Sigma_{f'}^{DL} = - Q^2_{f'} L/2$  
respectively. 
Contrary to the case of the $Z$ -boson decay,  
graph~(a) in Fig.~1 now yields a DL contribution 

\begin{equation}
\label{sigmaw}
P_W^{DL} =  Q^2_W L/2 ~  
\end{equation} 
and therefore 

\begin{eqnarray}
\label{quarks}
A_W^{(1)}   = A_W^{Born} 
\Big[ P_W^{DL} + \Sigma_f^{DL} + \Sigma_{f'}^{DL} \Big]=  \nonumber \\  
A_W^{Born} [Q^2_W -  Q^2_f - Q^2_{f'}] L/2 ~.   
\end{eqnarray} 
The electric 
charge conservation states that $Q_f - Q_{f'} = Q_W$. It allows 
 to rewrite  Eq.~(\ref{quarks}) in the form of Eq.~(\ref{woneloop}) .  

\section{Origin of the enhancement}
\label{ANAT}

Both the Sudakov suppression of Eq.~(\ref{zdecay}) and the enhancement of 
 Eq.~(\ref{wdecay}) have a very simple origin in terms of the algebra of 
the gauge group generators. We explain it, first 
using the Feynman gauge and then proceed to the case of the  
Coulomb gauge.  To begin with, let us remind that all  
electroweak boson fields $A,~Z,~W$ are linear combinations of the 
gauge isoscalar field $B$ and isovector fields $A_i ~(i = 1,2,3)$:
 
\begin{eqnarray}
\label{az}
W^{\pm} = \frac{A_1 \pm \imath A_2}{\sqrt{2}}, 
~Z = - B \sin \theta_W + A_3 \cos \theta_W,  \nonumber \\ 
~A =  B \cos \theta_W  + A_3 \sin \theta_W   
\end{eqnarray}   
where $\theta_W$ is the Weinberg angle. 
Therefore, propagator $D_{\mu \nu}^{(B)}$ of the isocalar field 
$B$ and propagator $D_{\mu \nu}^{(3)}$ of the isovector field $A_3$ 
can be expressed in terms of the photon propagator  
$D_{\mu \nu}^{(A)}$  and the $Z$ -boson propagator $D_{\mu \nu}^{(Z)}$. 
As DL contributions at the energies $\leq M$ 
come from the photon propagators only,  

\begin{equation}
\label{dlprop}
D_{\mu \nu}^{(B)} \approx  D_{\mu \nu}^{(A)} \cos^2 \theta_W ~,
~~~~~~~~D_{\mu \nu}^{(3)} \approx D_{\mu \nu}^{(A)} \sin^2 \theta_W ~.  
\end{equation}

The $SU(2)\,U(1)$ -parts of the 
vertexes of the interaction of fields $A_i$ and $B$ with fermions 
 are $g t_i A_i$ and  $g'(Y/2)  B$, with $t_i$ being the $SU(2)$ -generators 
$(i = 1,2,3)$,  
$Y$ being the fermion hypercharge and $g, g'$ being the electroweak 
couplings ($g' = g \tan \theta_W$). Now let us choose the Feynman 
gauge for the photon propagator  and consider again the 
one-loop graph in Fig.~1. We can regard both the external and virtual 
electroweak bosons in this Fig. as consisting of the gauge fields 
$B, A_i$. In order to discuss both the $Z$ and the $W$ -decays at the same 
time,  
let us suppose that the external field $X_r$ consists of 
the isoscalar and isovector fields:  $X_r = p B  + q A_r$ where 
$r = 1,2,3$ ; $p$ and $q$ are some scalar factors.   
Then the Born amplitude can be written as $p + q t_r$ and 
 the graph in Fig.~1 yields the following DL contribution:   

\begin{eqnarray}
\label{dloneloop}
A^{(1)}_r = - \Big[ g^2 \sin^2 \theta_W  (t_3 (p  + q t_r) t_3) + \nonumber \\ 
{g'}^2  \cos^2 \theta_W \frac{Y^2}{4} 
(p B + q t_r)\Big] (1/16 \pi^2) \ln^2(-M^2/\mu^2) ~.
\end{eqnarray}

Therefore, 

\begin{equation}
\label{t12}
A^{(1)}_r = -  A^{(Born)}_{r} \frac{\big(\xi_r + Y^2 \big)}{4} L 
\end{equation}
where $\xi_3 = 1,~\xi_{1,2} = -1$

Eq.~(\ref{t12}) 
leads to the suppression when $r = 3$ and to the 
enhancement when $r = 1,2$ provided $Y^2 < 1$. 
The exponentiation of DL contributions in terms of the gauge fields 
$B, A_3$ can be proved easily, e.g. by 
composing the Infrared Evolution 
Equations for amplitudes 
$A_r$ in the same way as it was done in Ref.~\cite{et}  
for calculating the DL asymptotics of the 
form factors of electron (Abelian gauge group) and quark 
(non-Abelian gauge group).      
Eq.~(\ref{t12}) reads that the first-loop DL contributions from 
exchange of the virtual isoscalar boson (the contributions $\sim Y^2$)   
always lead to the suppression whereas the exchanges of the isovector 
bosons lead either to the suppression or to the enhancement due to 
non-commutativity of the generators of the gauge group.  

In order to obtain this result 
when the Coulomb gauge is used, it is enough to notice  
that the only difference between DL contributions of the graphs in Fig.~2 
compared to Eqs.~(\ref{sigmaw},\ref{quarks}) 
is in $SU(2)$ -group factors. After accounting for them, one  
reproduces  Eq.~(\ref{t12}).

\section{Summary and outlook}
\label{SUM}        

We have considered the decays of the massive 
electroweak bosons into fermions in the 
DL approximation. We  
have shown that when the EW bosons are 
charged, accounting for the DL radiative corrections 
leads to  exponentiation of the first-loop 
contribution, however with the positive exponent,   
i.e. instead of the 
Sudakov suppression there is the enhancement for such decays. 
On the other hand, accounting for the soft photon emision accompanying 
such decays leads to the falling exponenentials. Therefore, these DL 
contributions cancel each other in expressions for inclusive 
cross-sections.     
 The results we obtained can be useful for analyses of $1 \to 2$ 
processes in theories with spontaneously broken 
non-Abelian gauge groups which involve   
different mass scales for charged massive bosons. 
 
\section{Acknowledgement}

The work is supported by grant CERN/FIS/43652/2001.

\end{document}